\documentclass[twocolumn,showpacs,amsmath,amssymb,prl,footinbib,floatfix,superscriptaddress]{revtex4}

\usepackage{graphicx}
\usepackage{dcolumn}
\usepackage{bm}

\usepackage{helvet}
\usepackage{amssymb}
\usepackage{amsmath}
\usepackage{amsfonts}
\usepackage{amssymb}
\usepackage{hyperref}
\usepackage{color}

\usepackage{xcolor}
\definecolor{red}{rgb}{0.7,0,0}
\definecolor{green}{rgb}{0.,0.35,0.}
\definecolor{blue}{rgb}{0.2,0.2,0.7} 
\definecolor{black}{rgb}{0.15,0.15,.15}

\def\barray{\begin{eqnarray}}
\def\earray{\end{eqnarray}}
\def\beq{\begin{equation}}
\def\eeq{\end{equation}}

\makeindex

\begin{document}

\title{Entanglement Entropies in Conformal Systems with Boundaries}

\date{\today}


\author{L. Taddia}
\affiliation{Dipartimento di Fisica e Astronomia dell'Universit\`a di Bologna and INFN, via Irnerio 46, 40126 Bologna, Italy}
\affiliation{Instituto de F\'{\i}sica Te\'orica  UAM/CSIC, Universidad Aut\'onoma de Madrid,  Cantoblanco 28049 , Madrid, Spain}
\author{J. C. Xavier}
\affiliation{Instituto de F\'{\i}sica, Universidade Federal de Uberl\^andia, Caixa Postal 593, 38400-902 Uberl\^andia, MG, Brazil}
\author{F. C. Alcaraz}
\affiliation{Instituto de F\'{\i}sica de S\~ao Carlos, Universidade de S\~ao Paulo, Caixa Postal 369, S\~ao Carlos, SP, Brazil}
\author{G. Sierra}
\affiliation{Instituto de F\'{\i}sica Te\'orica  UAM/CSIC, Universidad Aut\'onoma de Madrid,  Cantoblanco 28049 , Madrid, Spain}

\begin{abstract}
We study the  entanglement entropies in one-dimensional open critical systems, whose effective description is given by a conformal field theory with boundaries. We show that for pure-state systems formed by the ground state or by the excited states associated to  primary fields, the entanglement entropies have a finite-size behavior that depends on the correlation of the underlying field theory. The  analytical results  are checked numerically, finding excellent agreement for the quantum chains ruled by the theories with central charge $c=1/2$ and $c=1$. 
\end{abstract}













\maketitle
 
In recent years there has been a flurry of activity devoted to characterize  
quantum many body systems and quantum field theories  
using the concept of entanglement (see \cite{jpa} and references therein for a review).  
Under certain conditions,  the low energy states of local
Hamiltonians satisfy an entropic area law according to which
the entanglement entropy  (EE) of a subsystem is proportional to the area of its boundary \cite{ECP}.
In one spatial dimension,  violations to this area law appear if the system is gapless and described by a conformal field theory (CFT) \cite{BelavinPolyakovZamolodchikov1984, DiFrancescoMathieuSenechal1997}: in this case, the ground-state (GS) EE grows with the log of the subsystem size \cite{HLW,VLRK,CalabreseCardy}. On the other hand, it is well known that the dominant finite-size correction to the GS
energy of critical systems is related to the central charge $c$ \cite{finite}. A similar
correspondence also exists between the central charge and the
R\'enyi entanglement entropies (REE's) of the GS \cite{HLW,VLRK,CalabreseCardy}. In the same
vein, we would expect that the scaling dimensions of the primary
operators are related with the REE's  of the excited states (ES's), since
the finite mass gaps of  critical lattice Hamiltonians are also related with them \cite{finite}.
Indeed, this latter correspondence was proven recently for conformal
 systems with {\it periodic} boundary conditions \cite{AlcarazBerganzaSierra2011,BerganzaAlcarazSierra2012}.
The main result of these works is that the $n^{\rm th}$ REE is given in terms of  the $2n$-point correlator of the corresponding primary
field placed at special positions depending on the subsystem size. 

In this letter we shall generalize this result to {\it open} lattice Hamiltonians, effectively described by boundary conformal field theory (BCFT) \cite{Cardy1986}. BCFT has a wide range of applications to impurity problems \cite{Saleur1998}, string theory \cite{pol}, etc, which justifies the generalization pursued here. Moreover, the generalization is, as we shall see, non trivial, since BCFT has a very rich mathematical structure, which is worth to analyze from the entanglement point of view. Apart from these many reasons of interest, it is worth to cite that the problem is quite long-standing \cite{Zhou2006, AffleckLaflorencieSorensen2009}, and our work represents its final solution.
 
We shall start with some basic definitions. The  REE's
 are defined as follows: 
\begin{equation}
 S_n\equiv\frac{1}{1-n} \ln \mbox{Tr}_A \rho_A^n, 
\end{equation}
where $n$ is a positive real, $\rho_A$ is the reduced density matrix of  a subsystem $A$
and the trace is over the $A$'s Hilbert space. 
If $\rho_A$ is constructed from the GS of a CFT, one has 
  \cite{CalabreseCardy}:
\begin{equation}\label{CC}
 S_n^{CFT}(l,L)=c_n^\eta+\frac{c}{3\eta}\left(1+\frac{1}{n}\right)\ln\left[\frac{\eta L}{\pi}\sin\frac{\pi  l}{L}\right],
\end{equation}
where $l$ is the size of $A$, 
$L$ is the total-system size and $\eta=1,2$ for periodic/free boundary conditions (PBC/FBC); 
$c$ is the central charge and $c_n^\eta$ is a constant that depends on the BC's 
\cite{DiFrancescoMathieuSenechal1997}. For open systems, $l$ is measured from the left edge. 
Equation (\ref{CC}) may have significant corrections which carry useful information about the underlying CFT 
\cite{Laflorencie2006,Calabrese2010,XavierAlcaraz2012}. 
Different kind of corrections arise when, instead of considering the GS, one looks at the ES's \cite{AlcarazBerganzaSierra2011,BerganzaAlcarazSierra2012,DalmonteErcolessiTaddia2012,EloyXavier2012},
 or when the system satisfies general BC's preserving its conformal invariance (FBC are just one of them) 
 \cite{Zhou2006,AffleckLaflorencieSorensen2009}. In particular, the corrections in this last case have, up to now, 
 not have been derived analytically, and their knowledge would be a great advance, both from a practical and a 
conceptual point of view. In the following, we present the general CFT framework allowing the analytical computation of such corrections. 
These results are then verified  with DMRG calculations in two examples: 
the $c=1/2$ minimal CFT and the $c=1$ massless compactified free boson. 

Let us consider a 1D conformal system, defined on a strip of width $L$. 
Cardy \cite{Cardy1986} showed the existence of  
BC's, denoted as  $\{\tilde{\alpha}\}$, that preserve the conformal invariance, 
and such that the  partition function takes the form
\begin{equation}\label{Z}
 Z_{\tilde{\alpha}\tilde{\beta}}(q)=\sum_h\mathcal{N}^h_{\tilde{\alpha} \tilde{\beta}}\chi_h(q),
\end{equation}
where $q\equiv e^{-\pi\beta/L}$ (being $\beta$ the inverse temperature),   
$h$ are the conformal dimensions of the primary fields, and $\chi_h$ are  the associated  characters. 
The integers $\mathcal{N}^h_{\alpha\beta}$ are the fusion coefficients of the theory \cite{BelavinPolyakovZamolodchikov1984,DiFrancescoMathieuSenechal1997}. 
The derivation of the 
REE's of ES's, originating from the primary operator $\Upsilon(w)$ for open conformal systems
follows closely the one of periodic boundary conditions given in   \cite{AlcarazBerganzaSierra2011}. 
The REE's are given by 
\barray
 S_n(l,L) & = & S_n^{CFT}(l,L)+\frac{1}{1-n}\ln F^{(n)}_\Upsilon,  \label{ree} \\ 
 F^{(n)}_\Upsilon(x)  & = &  \frac{   {\rm Tr}_A  \, \rho^n_{A, \Upsilon}}{ {\rm Tr}_A  \, \rho^n_{A, GS}},  \nonumber 
\earray 
where $\rho_{A, \Upsilon}$ and $\rho_{A, GS}$ are the reduced density matrices obtained from the ES produced by the primary field $\Upsilon$ and from the GS, respectively.
The ratio $ F^{(n)}_\Upsilon$ can be computed using  path integral methods \cite{AlcarazBerganzaSierra2011}.
We explain below the main  steps of the derivation.  

Let us first split  the strip into the  subsystems  $A=[0,l], B = [l, L]$. 
This strip  is parameterized by the complex coordinate $w= \sigma +i\tau$, $\sigma \in[0,L]$, 
$\tau\in ( -\infty,\infty )$.  We make the conformal transformation 
\beq
\zeta = \frac{  \sin \frac{\pi ( w - l)}{2 L}}{ \sin \frac{\pi ( w + l)}{2 L}  }, 
\label{c1}
\eeq
which maps the strip into the unit disc $\mathbb D= \{\zeta, | \zeta| \leq 1 \}$. 
The intervals $A$ and $B$ are  mapped  into the segments $(-1, 0)$ 
and  $(0,1)$ respectively. The boundaries of the strip, $\sigma =0, L$ are mapped
into the boundary of the disc ($| \zeta|=1$),  and the infinite past ($w_\infty^-= - i \infty$)  and infinite future ($w^+_\infty= i \infty$)
are mapped into 
\beq
w_\infty^-  \rightarrow  \zeta_\infty^- = e^{ - i \pi x}, \quad 
w_\infty^+  \rightarrow \zeta_\infty^+ = e^{  i \pi x}, \quad x \equiv \frac{l}{L}.  
\label{c2}
\eeq
Next, we make $n$ copies of the unit disc and sew  them  along the cut  $(-1,0)$, obtaining the Riemann surface
${\cal R}_n$, which can also  be mapped into the unit disc by the conformal transformation
\beq
z= \zeta^{1/n} = \left(   \frac{  \sin \frac{\pi ( w - l)}{2 L}}{ \sin \frac{\pi ( w + l)}{2 L}  } \right)^{1/n}. 
\label{c3}
\eeq
The points $\zeta_\infty^\pm$ give rise to $2 n$  points on  ${\cal R}_n$ 
\beq
z_{k, n}^\pm = e^{ \frac{ i \pi}{n} ( \pm  x + 2 k)},  \quad k=0,1,  \dots, n-1, 
\label{c4}
\eeq
where the primary fields $\Upsilon$ and its conjugate $\Upsilon^\dagger$ are inserted.
Repeating the same steps as in \cite{AlcarazBerganzaSierra2011}, we arrive
at  an expression for Eq. (\ref{ree})  
\begin{equation}
 F_\Upsilon^{(n)}(x)=\frac{e^{i2\pi(n-1)h}}{n^{2nh}}\frac{\left<\prod_{k=0}^{n-1}\Upsilon(z_{n,k}^-)\Upsilon^\dagger(z_{n,k}^+)\right>}{\left<\Upsilon(z_{1,0}^-)\Upsilon^\dagger(z_{1,0}^+)\right>^n}, 
 \label{fcom}
 \end{equation}
where we have assumed that $\Upsilon$ is a chiral primary field with conformal dimension $h$.
A similar formula holds for non chiral fields. The correlators in Eq.(\ref{fcom}) are computed 
on the unit disk for  fields inserted at its boundary. These primary fields change the boundary
conditions $a$ and $b$ on each edge of  the strip: $\Upsilon$ changes them from $a$ to $b$, while
$\Upsilon^\dagger$ from $b$ to $a$. Hence the numerator of Eq. (\ref{fcom})
is proportional to the partition function of a disc with $2n$ segments where the boundary
conditions $a$ and $b$ alternate \cite{DiFrancescoMathieuSenechal1997}. 
Equation (\ref{fcom}) 
constitutes the main result of the work, which we shall verify below  for two models. 

{\it The $c=1/2$ minimal CFT.}  This CFT contains three primary fields: 
 the identity $\mathbb{I}$ ($h_{\mathbb{I}}=0$), a Majorana fermion $\chi$ ($h_\chi=1/2)$ 
 and  a spin  field $\sigma$ ($h_\sigma=1/16$), whose fusion rules are
\begin{equation}\label{fusion_ising}
 \begin{split}
  \mathcal{N}^h_{00}=\mathcal{N}^h_{\frac{1}{2}\frac{1}{2}}=\delta^h_0,\;\;\mathcal{N}^h_{0\frac{1}{2}}=\delta^h_{\frac{1}{2}}, \\
  \mathcal{N}^h_{0\frac{1}{16}}=\mathcal{N}^h_{\frac{1}{16}\frac{1}{2}}=\delta^h_\frac{1}{16},\;\;\mathcal{N}^h_{\frac{1}{16}\frac{1}{16}}=\delta^h_0+\delta^h_{\frac{1}{2}}. 
 \end{split}
\end{equation}
This CFT describes the long distance properties of the critical Ising model in a transverse field
whose lattice Hamiltonian is 
%
\begin{equation}\label{Ising}
 H_I=-\frac{1}{2}\sum_{j=1}^{L-1}\sigma_j^x\sigma_{j+1}^x-\frac{1}{2}\sum_{j=1}^{L}\sigma_j^z, 
\end{equation}
where $\sigma^{x,y,z}$ are  the Pauli matrices. 
The correspondence between the conformal BC's $\{\tilde{\alpha}\}$ and the lattice BC's 
is the following:  $\tilde{0}$ ($\tilde{\frac{1}{2}}$), corresponds 
to fix ${\sigma}^x_{1,  L}$ to $+1$ (-1), while $\tilde{\frac{1}{16}}$ corresponds to free BC's \cite{Cardy1986,Zhou2006}. 
For simplicity,  we shall denote  $\tilde{0}, \tilde{\frac{1}{2}}, \tilde{\frac{1}{16}}$ by $+, -, F$.


We begin by considering the REE's of the GS  with $FF$BC's (the two  $F$'s refer to free BC's
on both edges).  
The spin Hamiltonian (\ref{Ising}) can be  mapped into  a free fermion Hamiltonian  \cite{LiebSchultzMattis1961},
which we use to compute, by using the method of \cite{Peschel2003}, the von Neumann entropy, i.e., the $n=1$ REE,  for chains
with lengths 60-180 in multiples of 20 (this will be the sizes considered in the rest of the work). 
 Using Eq. (\ref{CC}), and  
 finite-size scaling (FSS) techniques, we obtain  the asymptotic values  $c=0.499$ and  $c_1^{\eta=2}=0.241$.
 The value of $c$ agrees to high precision with the central charge of the critical Ising model. 
 Moreover, the value  of $c_1^{\eta=2}$ is very close to $c_1^{\eta=1}/2$, that can be obtained with PBC.
 The relation $c_1^{\eta=2}-c_1^{\eta=1}/2=\ln g$ 
 \cite{CalabreseCardy,Zhou2006}, is satisfied in this case, because the boundary entropy (BE) $\ln g$ 
 vanishes for $FF$ BC's  \cite{AffleckLudwig1991}. Moreover, as a different reliability check, we verified that the REE's in this case are exactly one half of the ones of an XX spin-$1/2$ chain with $FF$ BC's (see below), according to reference \cite{IgloiJuhasz2008}.
 
\begin{figure}[t]
 \begin{minipage}{0.5\textwidth}
  \includegraphics[width=0.49\textwidth]{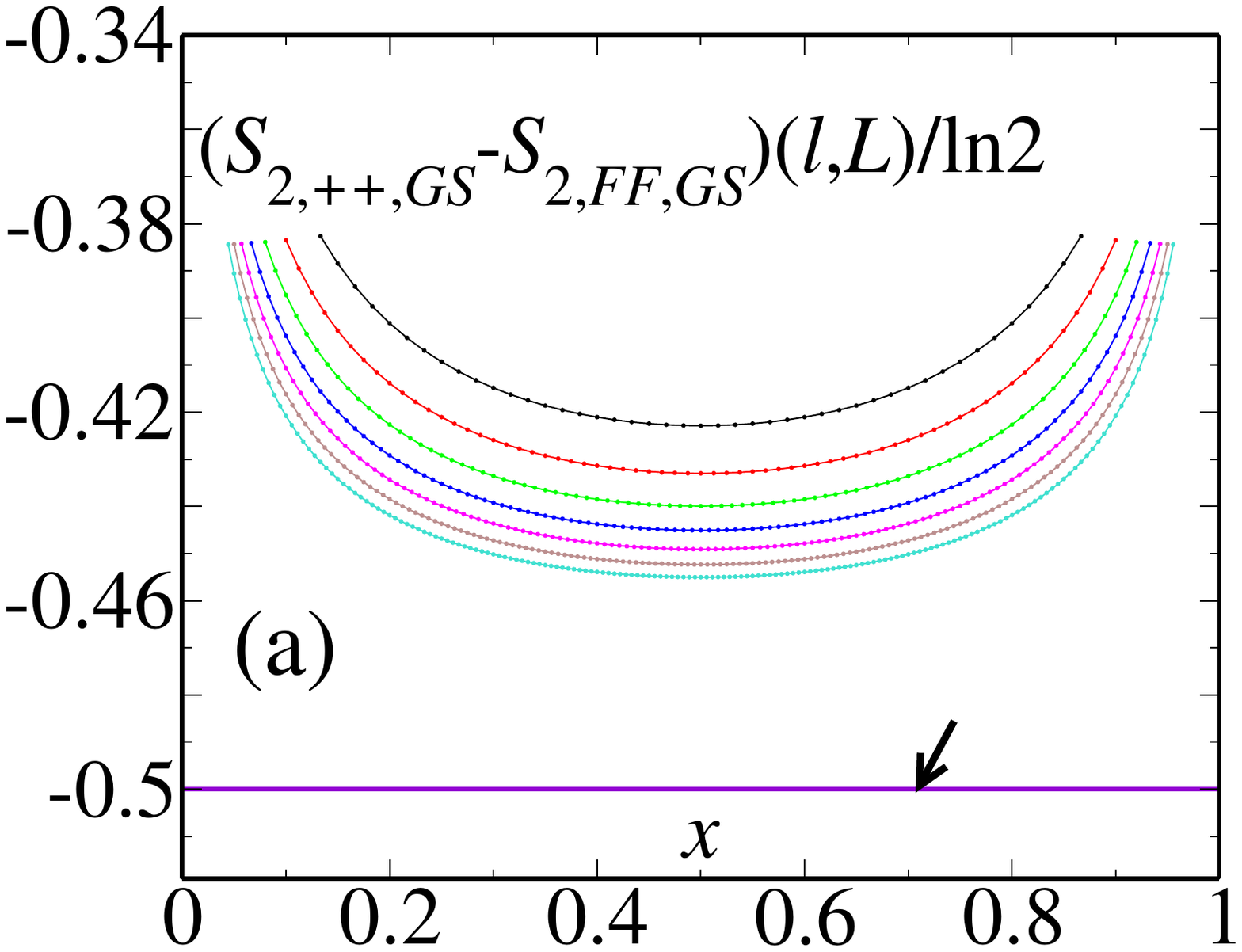}
  \includegraphics[width=0.49\textwidth]{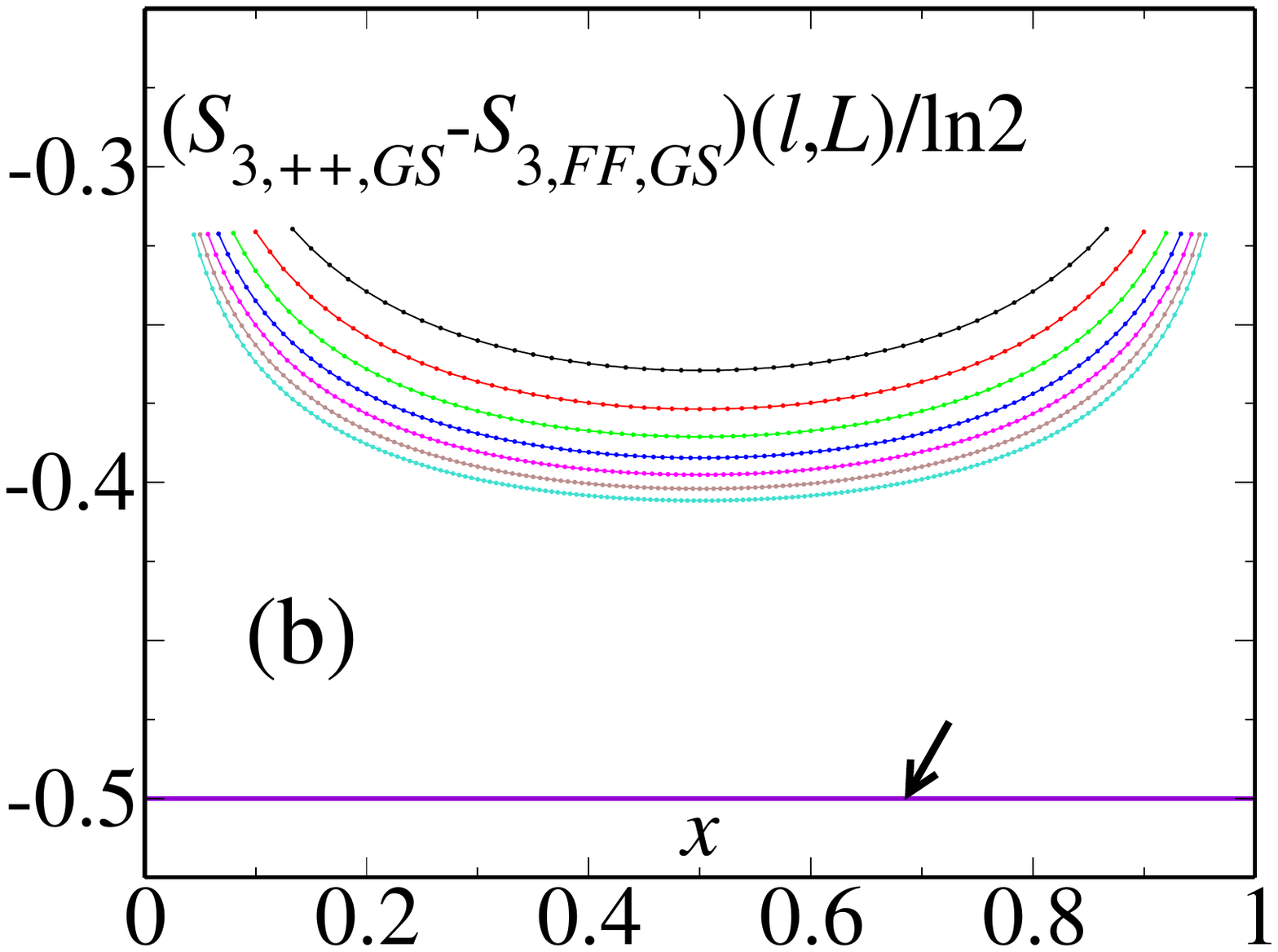}
  \includegraphics[width=0.49\textwidth]{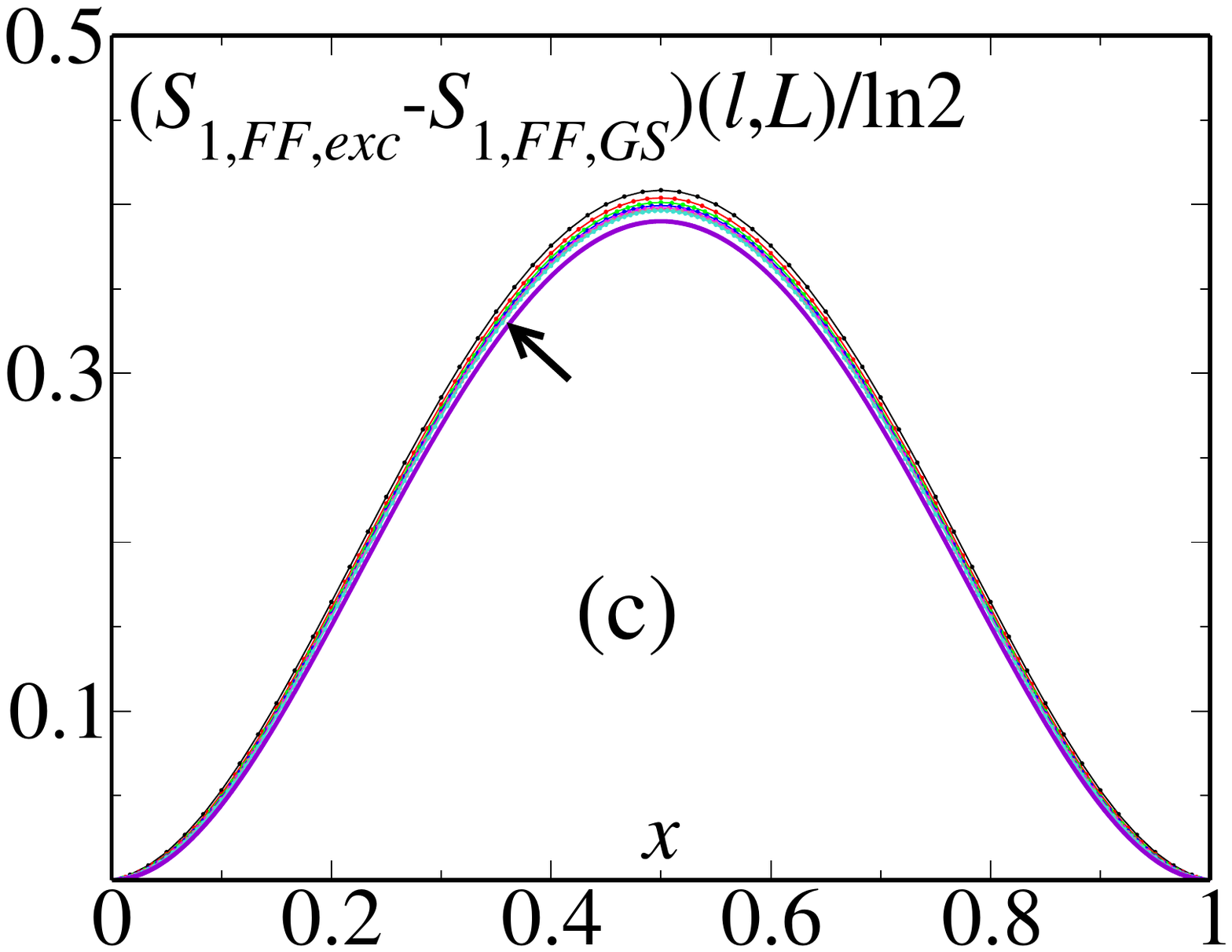}
  \includegraphics[width=0.49\textwidth]{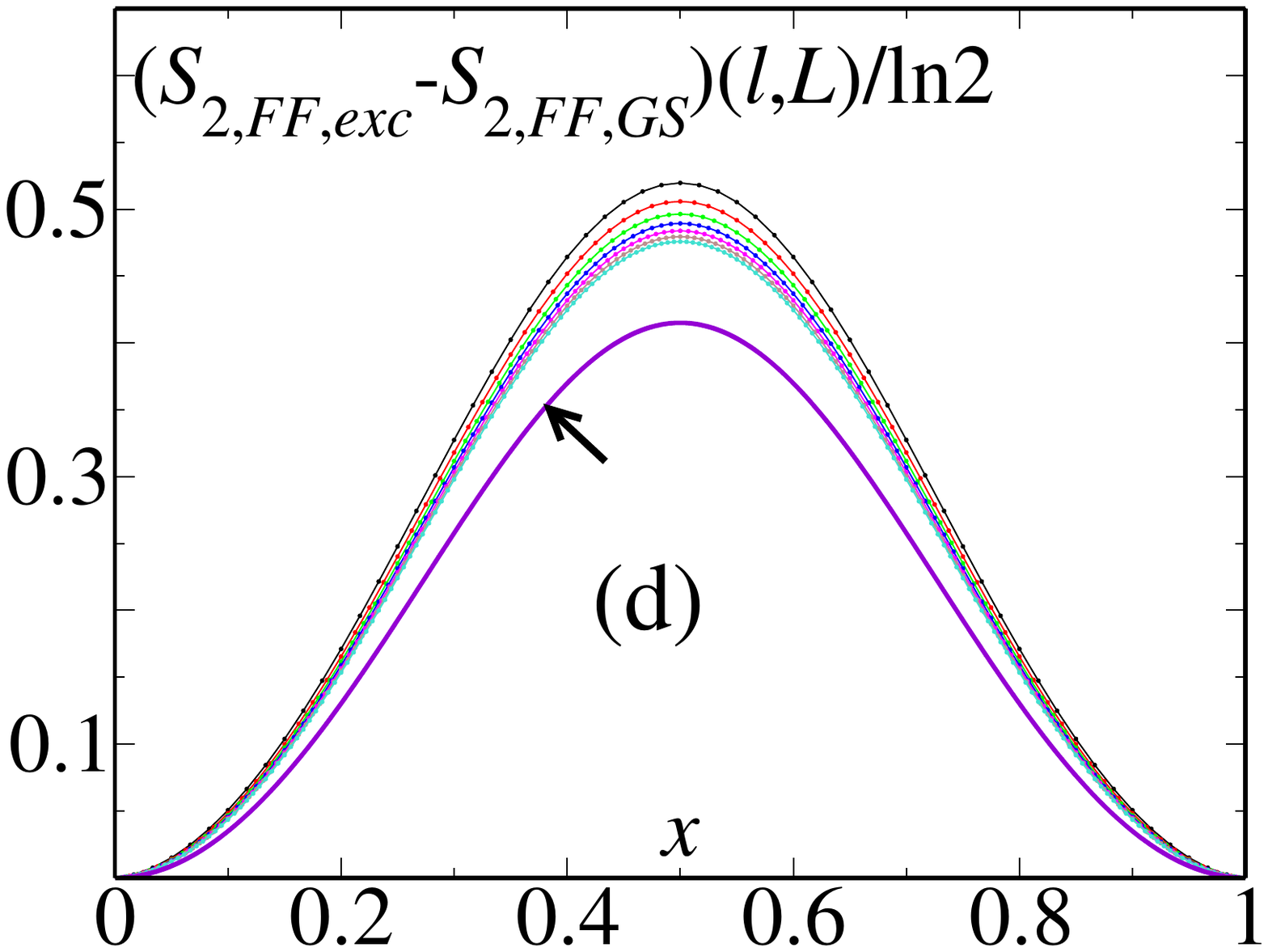}
  \includegraphics[width=0.49\textwidth]{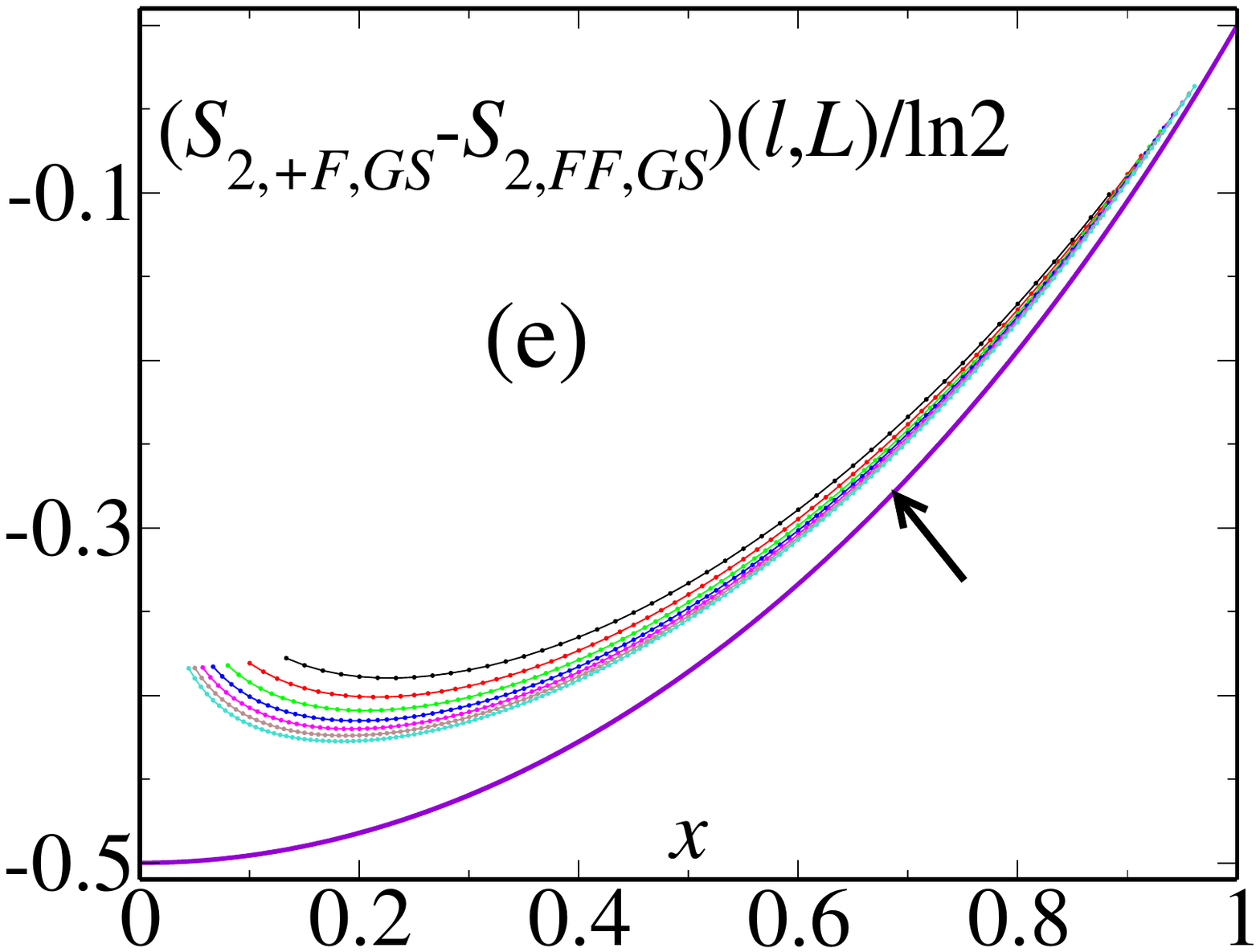}
  \includegraphics[width=0.49\textwidth]{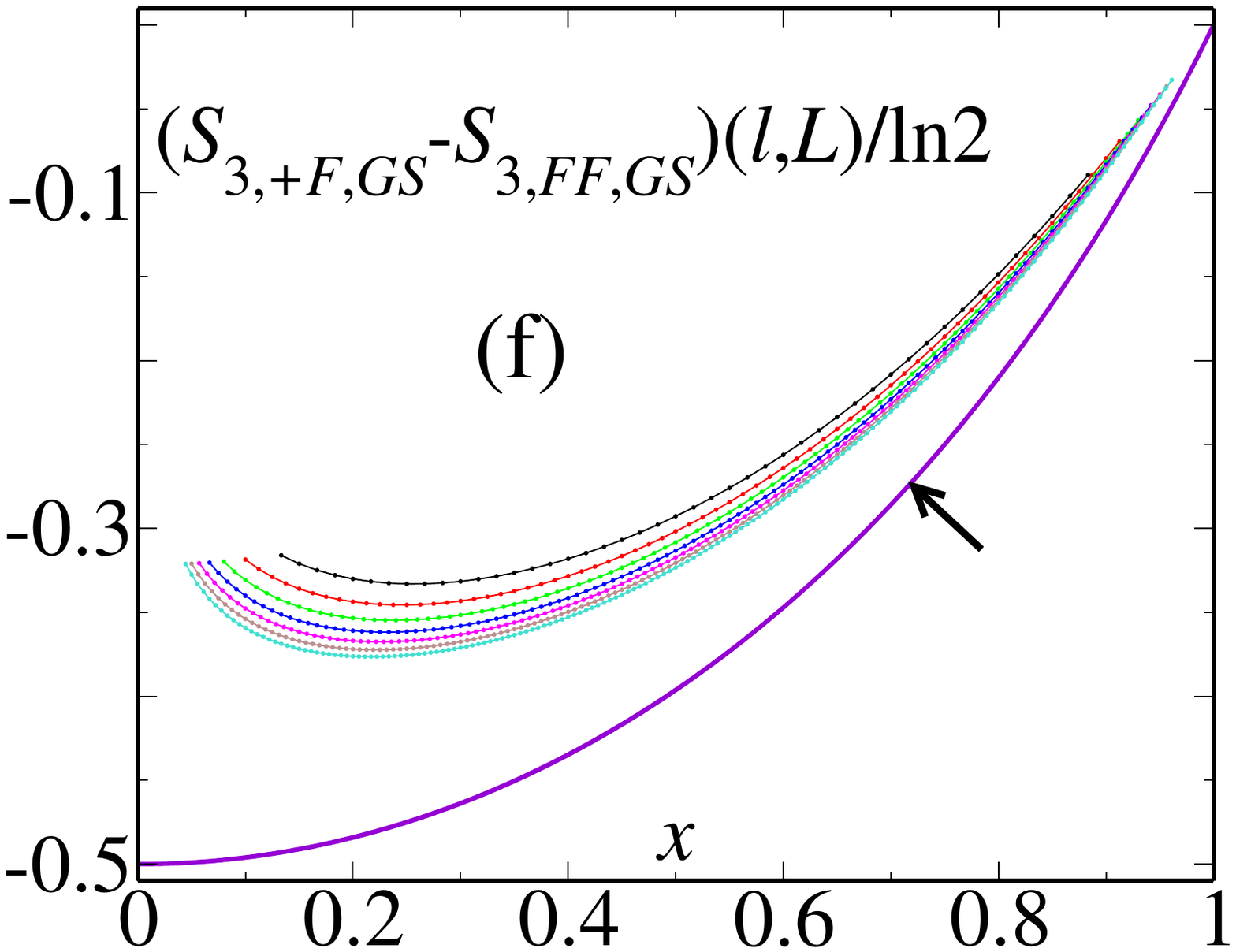}
  \includegraphics[width=0.49\textwidth]{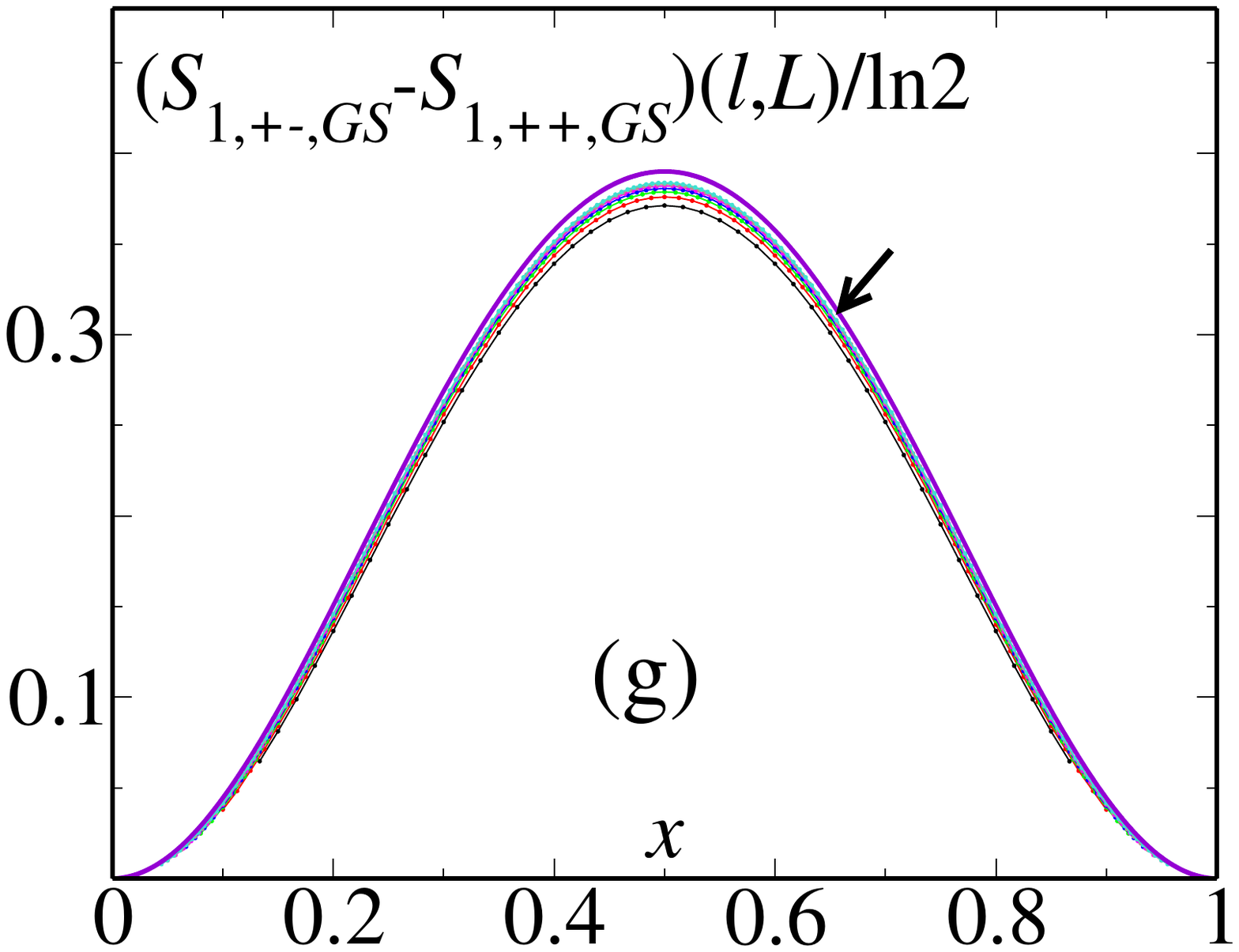}
  \includegraphics[width=0.49\textwidth]{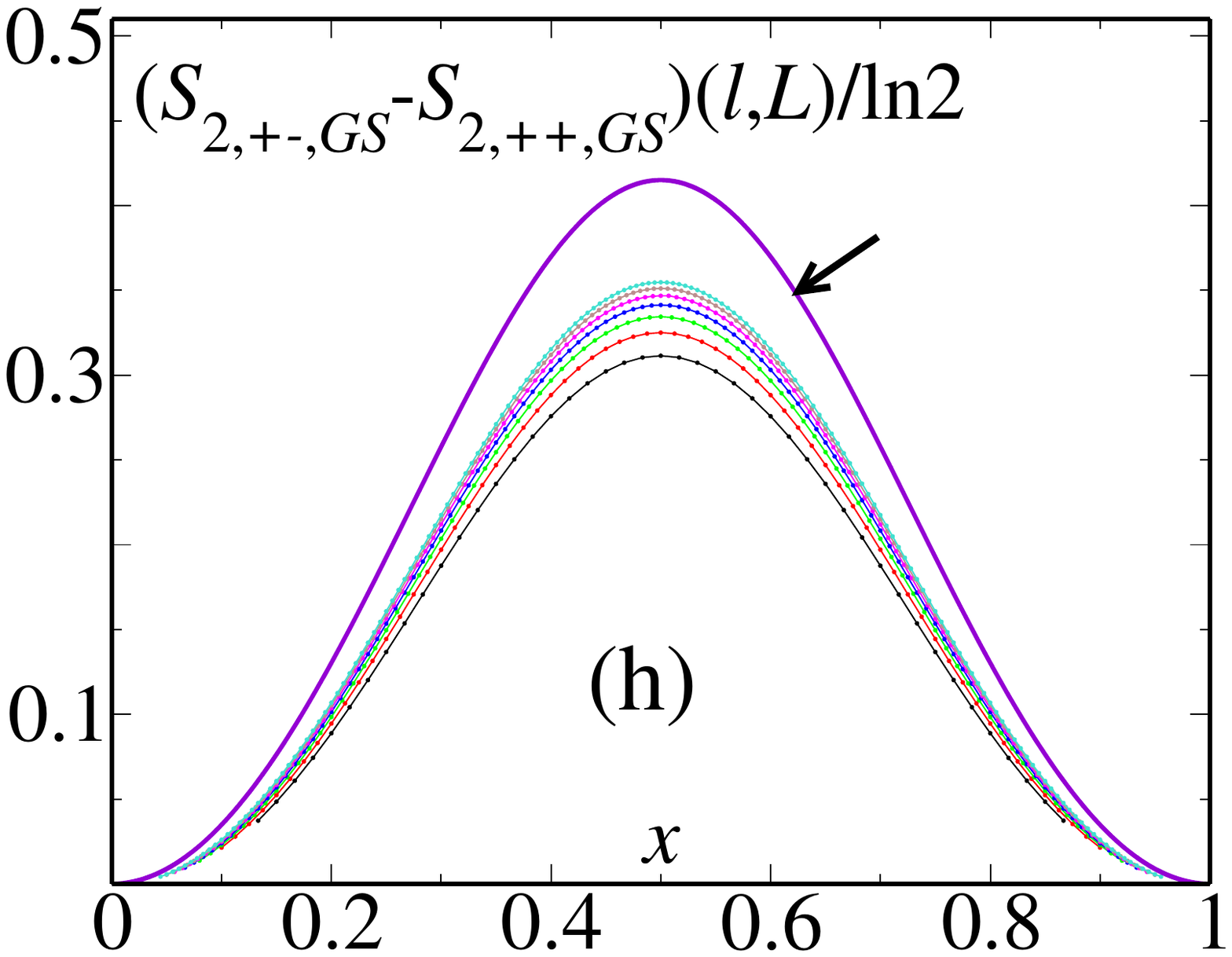}
  \includegraphics[width=0.49\textwidth]{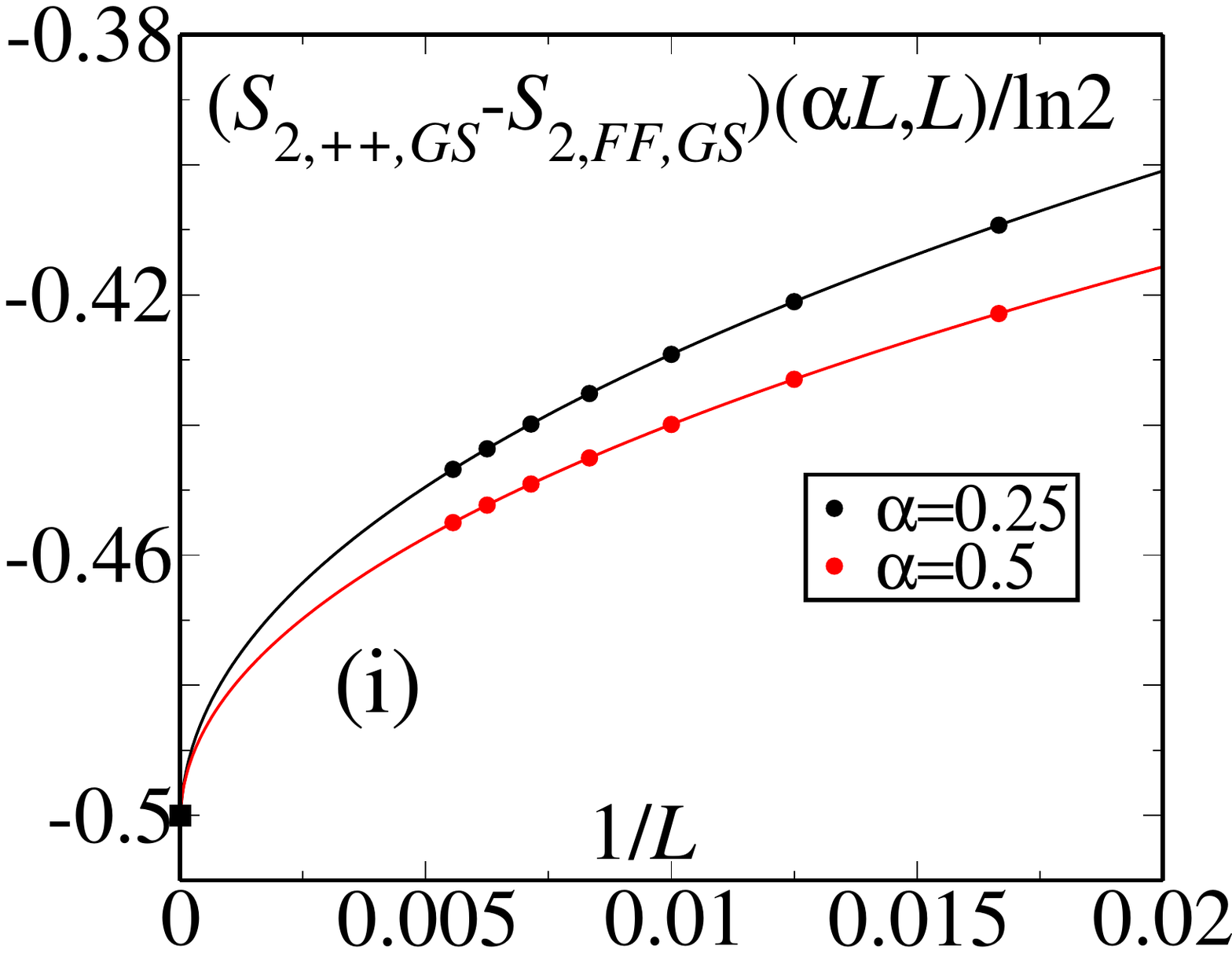}
  \includegraphics[width=0.49\textwidth]{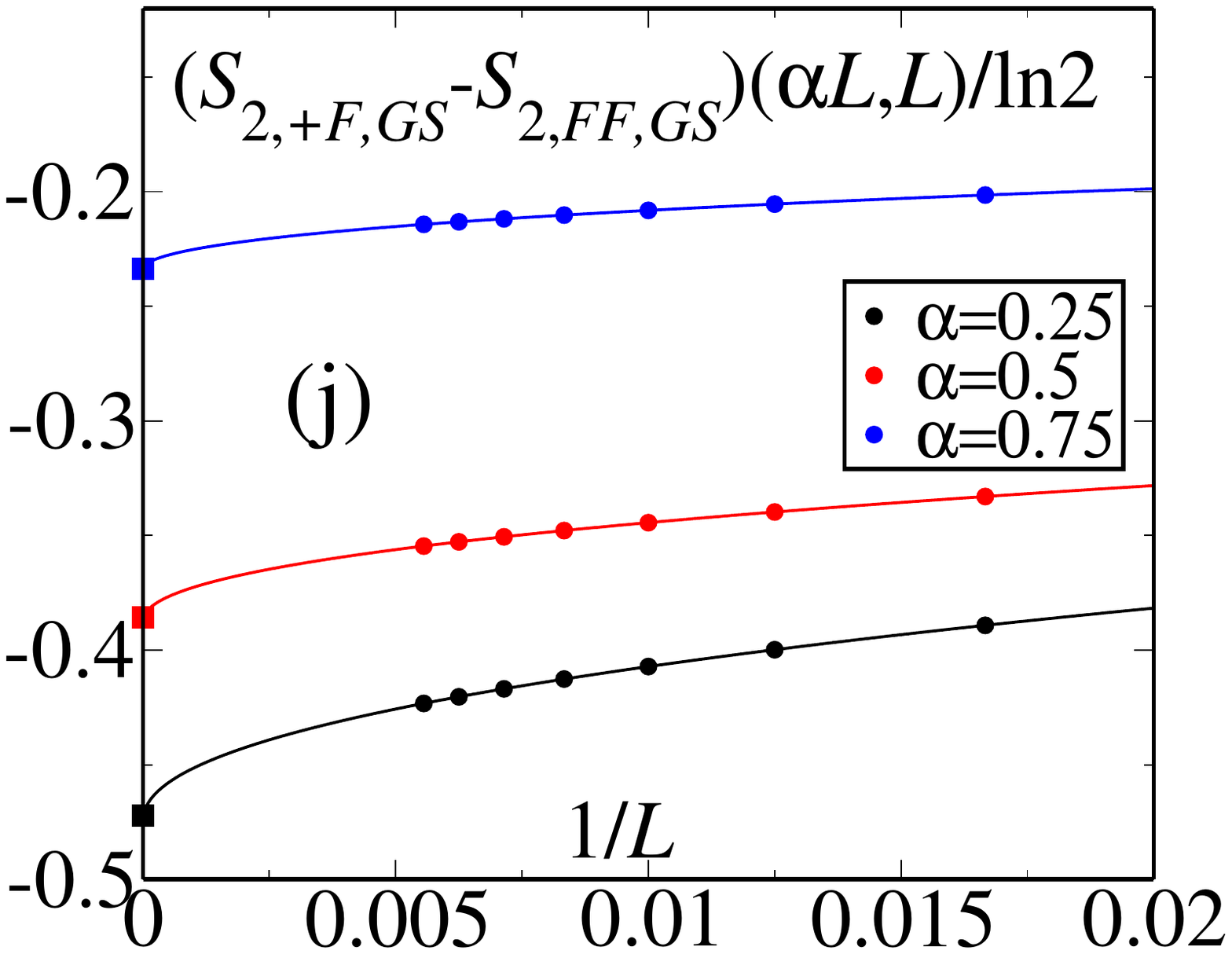}
 \end{minipage}
 \caption{REE's in the 1D critical Ising model. Panels (a)-(h): black to turquoise (dotted lines): $L=60$ to 180 
 numerical data; purple (solid line) + arrow: CFT predictions (the BE's are added to the CFT formulas when necessary); (i), (j): FSS relative to panels (a) and (e) (best fits use the 5-parameters formula $y=a_0+a_1x^{a_2}+a_3x^{a_4}$; dots are numerical data, solid lines best fits and squares the CFT predictions).}\label{Ising_fig}
\end{figure}

We next  study the $++$BC case (that is equivalent to  $--$BC). 
The fusion rule  ${\cal N}_{00}^h = \delta_0^h$, lead us to consider only the case where $h=0$, for which no corrections arise (apart, as we shall see, from constant BE contributions). 
We compute the REE's for the Hamiltonian (\ref{Ising}) using the DMRG method \cite{White} with  up to 800 states per 
block,  and 3 sweeps, which yields  a truncation error of $10^{-12}$ or less.
 The Figs.  \ref{Ising_fig}(a), (b) display  the results for $n=2,\,3$ REE's: 
the data progressively flattens to the theoretical  value $-\frac{1}{2}\ln 2$ \cite{Cardy1989,Zhou2006}.
The convergence to the CFT predictions is of order $10^{-3}$ or less, as 
 confirmed by a FSS analysis (Fig. \ref{Ising_fig}(i)). 
This behavior can be ascribed to the presence of a slowly $L$-depending finite-size correction \cite{IgloiJuhasz2008}, previously observed, for the Ising model, in \cite{BerganzaAlcarazSierra2012}.

We now consider  the $FF$BC case. The fusion rule $\mathcal{N}^\chi_{\sigma, \sigma}=1$   
implies  that the first ES is generated  by the primary field $\chi$.  The associated $F$ function
is given by   $F_\chi^{(n)}=\sqrt{F_{i\partial\phi}^{(n)}}$, where $\phi$ is a massless free boson with $c=1$ \cite{BerganzaAlcarazSierra2012}. 
 Quite interestingly, $F_{i\partial\phi}^{(n)}$ has a general expression valid for any value of $n >   0$ \cite{EsslerLauchliCalabrese2012}
\begin{equation}\label{ELC}
 F^{(n)}_{i\partial\phi}(x)=\left\{\left[\frac{2\sin(\pi x)}{n}\right]^n\frac{\Gamma\left(\frac{1+n+n\csc(\pi x)}{2}\right)}{\Gamma\left(\frac{1-n+n\csc(\pi x)}{2}\right)}\right\}^2. 
\end{equation}
The Figs.  \ref{Ising_fig}(c) ,(d)  show the convergence of  the numerical results to 
the CFT prediction obtained with (\ref{ELC}), for $n=1,\,2$. 

We then consider the $+F$BC case. The fusion rule 
$\mathcal{N}^\sigma_{\mathbb{I}, \sigma}=1$ implies that 
the spectrum contains just the conformal tower of the $\sigma$ field, 
so the $F$ function must be
 $F^{(n)}_\sigma$. The nontrivial fusion rule of the field $\sigma$ yields
 different chiral correlators, which were computed for general $n$ in  \cite{ArdonneSierra2010}. 
 In our case though, only a combination of them yields the appropriate $F$ functions, which for $n=2,3$ 
 are given by 
%
\begin{equation}
 F_{\sigma}^{(2)}(x) = 
  \cos\frac{\pi x}{4}, \quad 
  F_{\sigma}^{(3)}(x) =  
  \cos \frac{\pi x}{3}. 
\end{equation}
%
The Figs. \ref{Ising_fig}(e), (f) (also see Fig. \ref{Ising_fig}(j)) display  the DMRG results and the CFT predictions: 
the agreement, for $L\rightarrow\infty$ is excellent, up to the constant term  $-\frac{1}{2}\ln 2$.

Finally, we consider the $+-$BC case (equivalent to the $-+$BC case). 
The spectrum contains just the conformal tower of the $\chi$ operator, and therefore the REE's of the GS shall be 
the same of the ones of the first ES in the $FF$BC case (see
Figs. \ref{Ising_fig}(g), (h)).
To conclude, we have shown that the CFT predictions for the Ising model with all possible
conformal BC's agree with the numerical results, up to the constant contribution due to the BE. 

{\it The $c=1$ compactified free boson.} 
We shall next consider a massless free boson with a compactification radius $R=1$. 
This CFT is rational, meaning that the chiral symmetry is enhanced in such a way that the number
of primary fields is finite and have conformal dimensions $h=0, 1/8, 1/2, 1/8$ \cite{DiFrancescoMathieuSenechal1997}. 
The corresponding conformal characters are given by setting $\lambda=0,\ 1,\ 2,\ 3\mod 4$ ($h_\lambda=\lambda^2/8$) in

\begin{equation}\label{Kharacters}
 K_\lambda(q)\equiv\frac{1}{\eta(q)}\sum_{n\in\mathbb{Z}}q^{\frac{1}{8}(4n+\lambda)^2},
\end{equation}
where $\eta(q)$ is the Dedekind function. The conformal BC's are  Dirichlet ($D$) and Neumann ($N$), and the corresponding boundary entropies are $0$ and $-\frac{1}{2}\ln2$ \cite{Saleur1998}.

\begin{figure}[t]
 \begin{minipage}{0.5\textwidth}
  \includegraphics[width=0.49\textwidth]{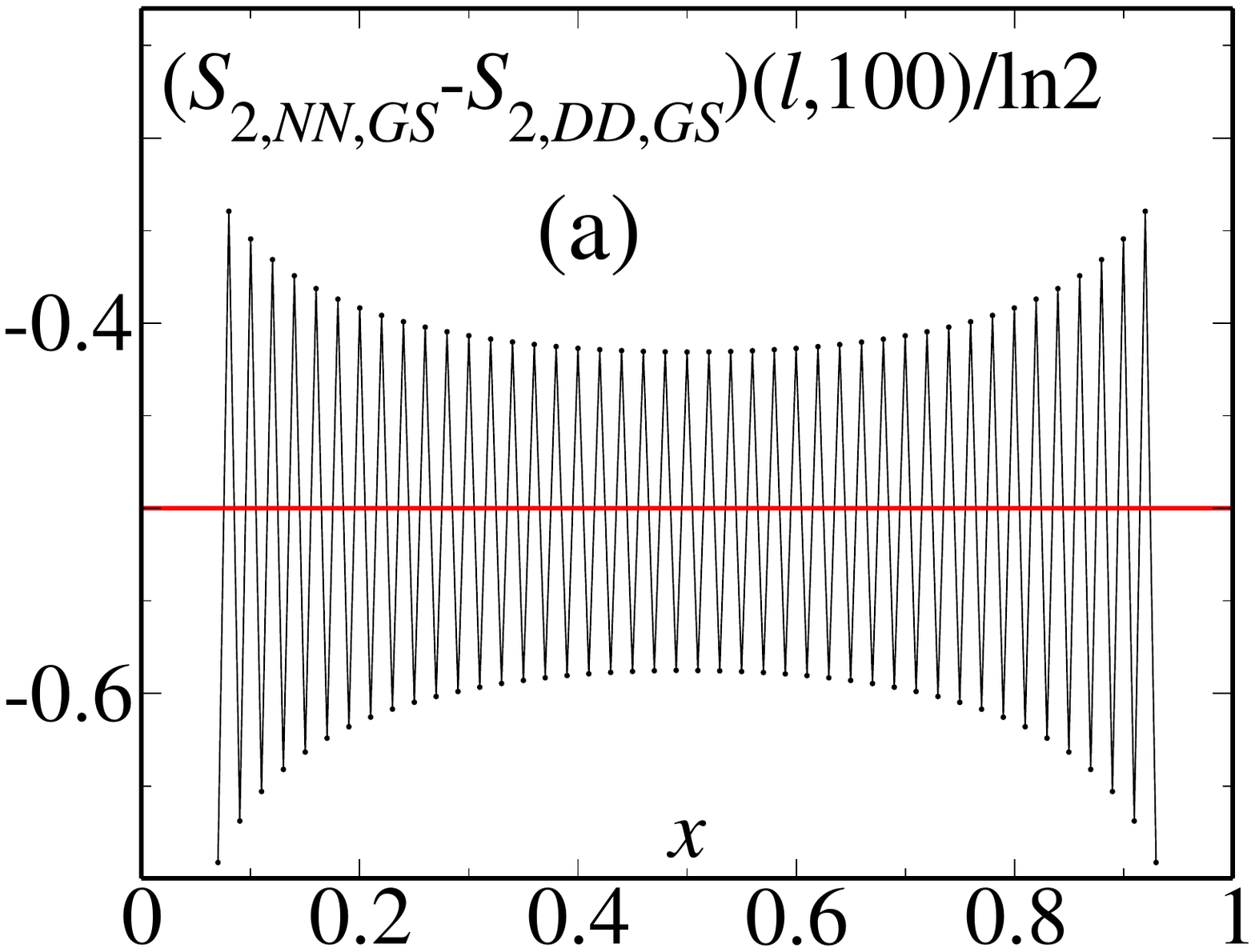}
  \includegraphics[width=0.49\textwidth]{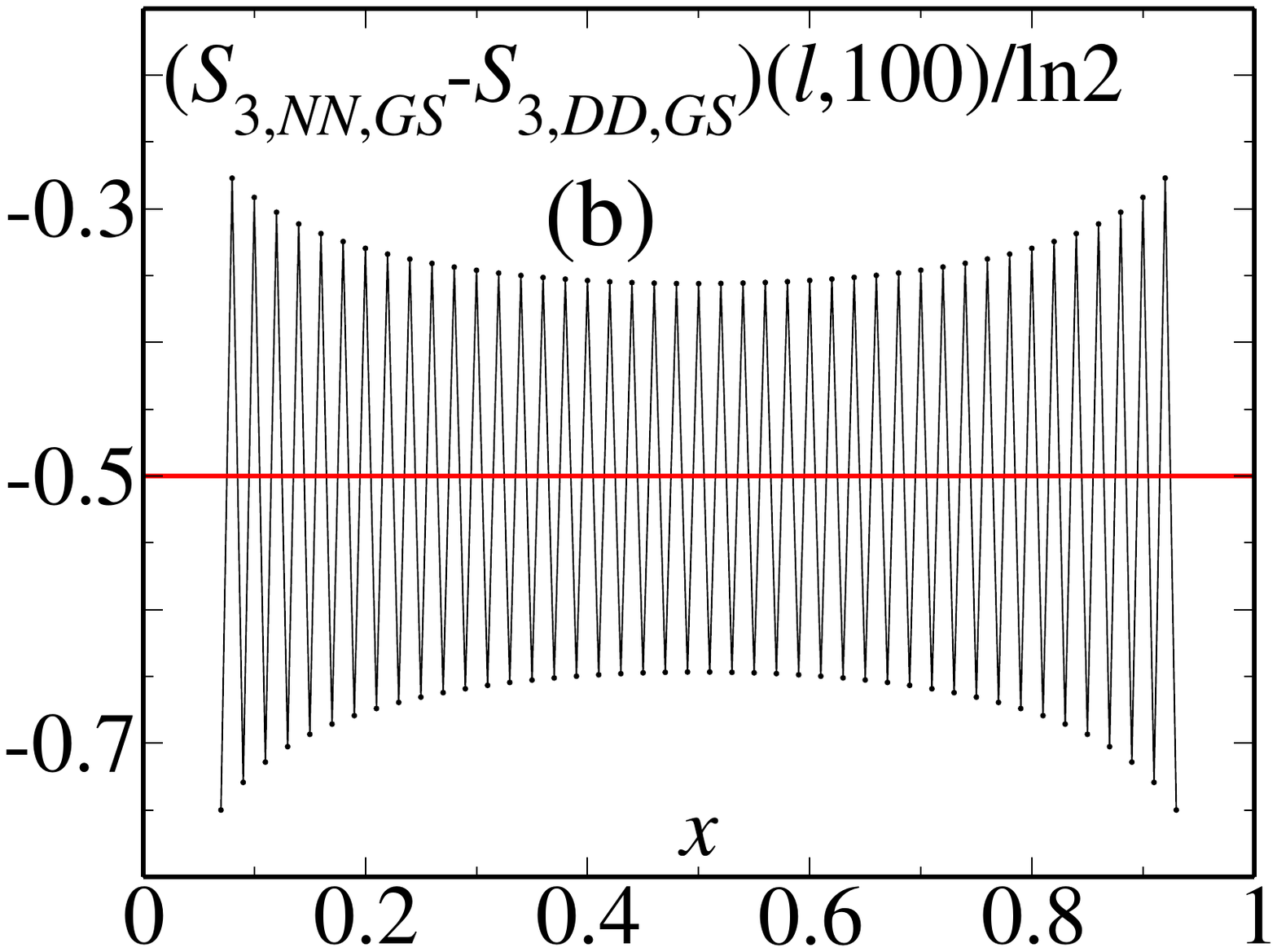}
  \includegraphics[width=0.49\textwidth]{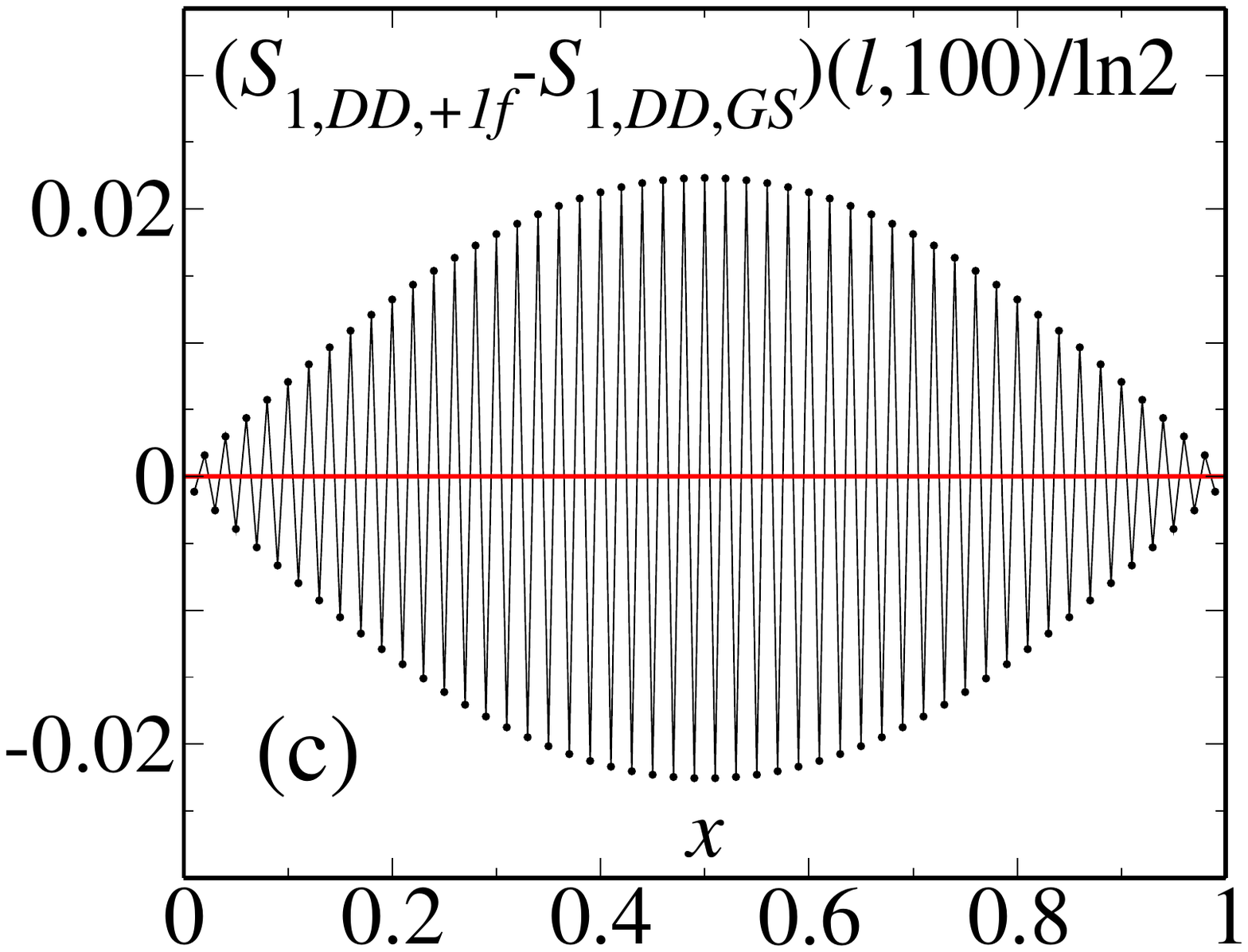}
  \includegraphics[width=0.49\textwidth]{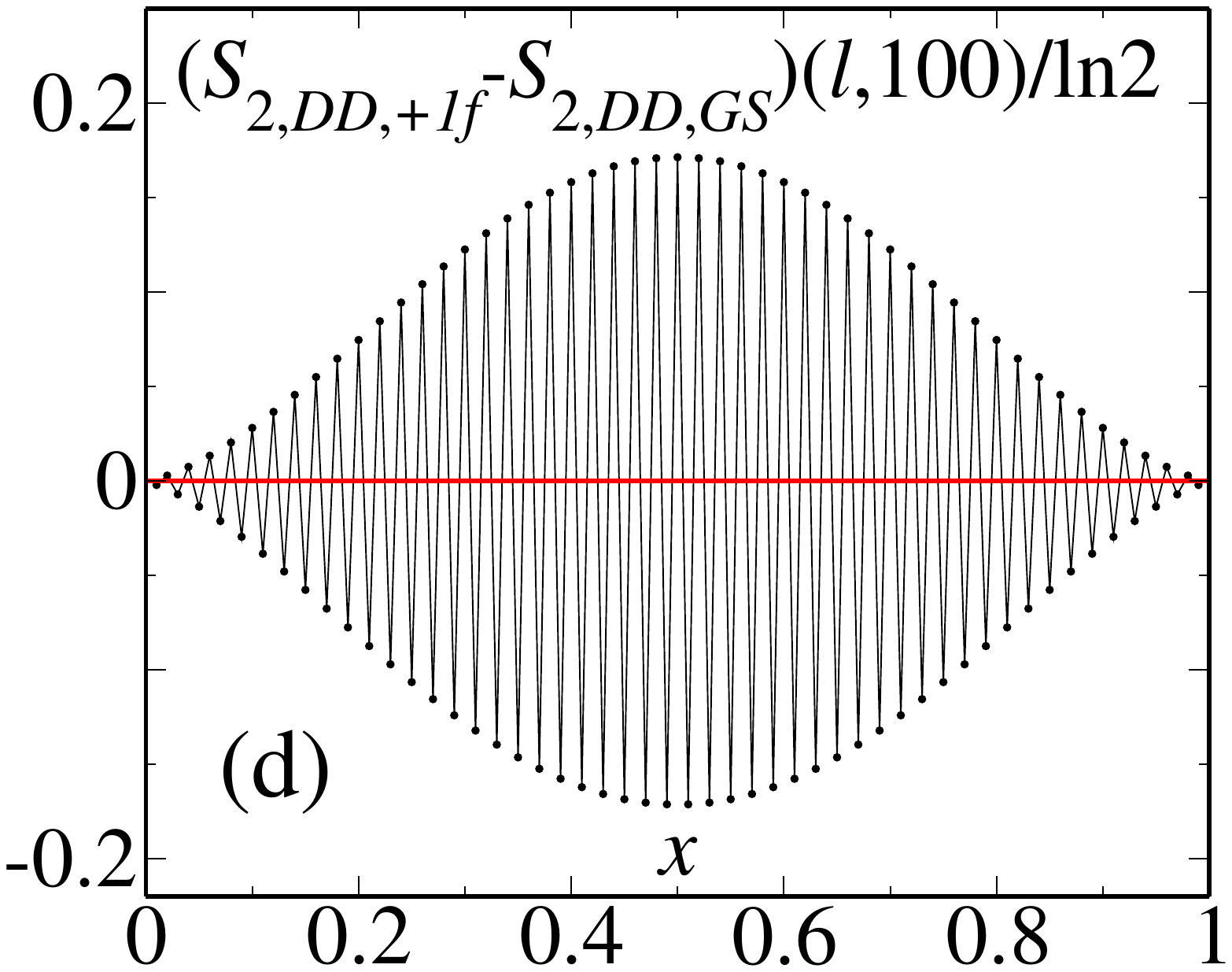}
  \includegraphics[width=0.49\textwidth]{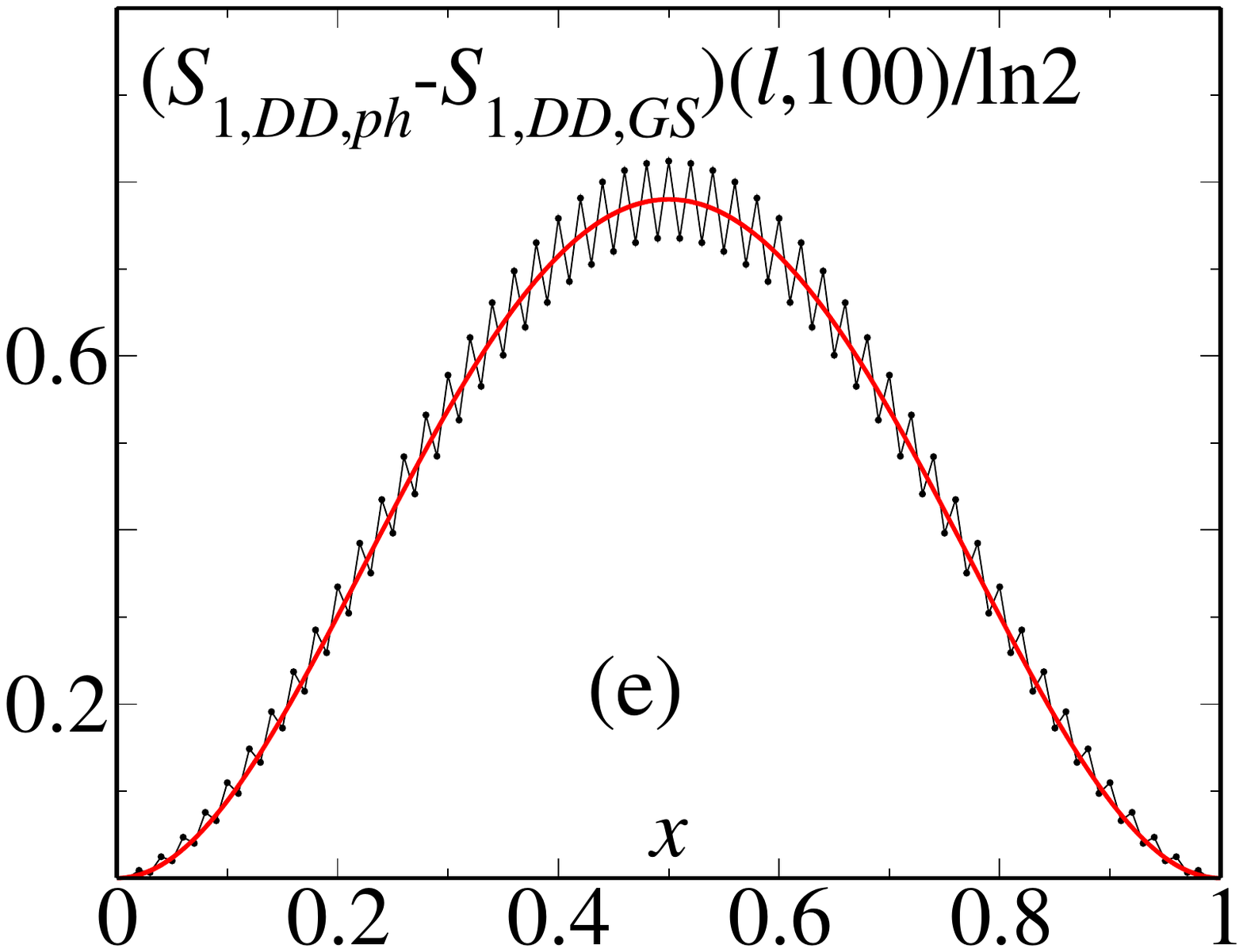}
  \includegraphics[width=0.49\textwidth]{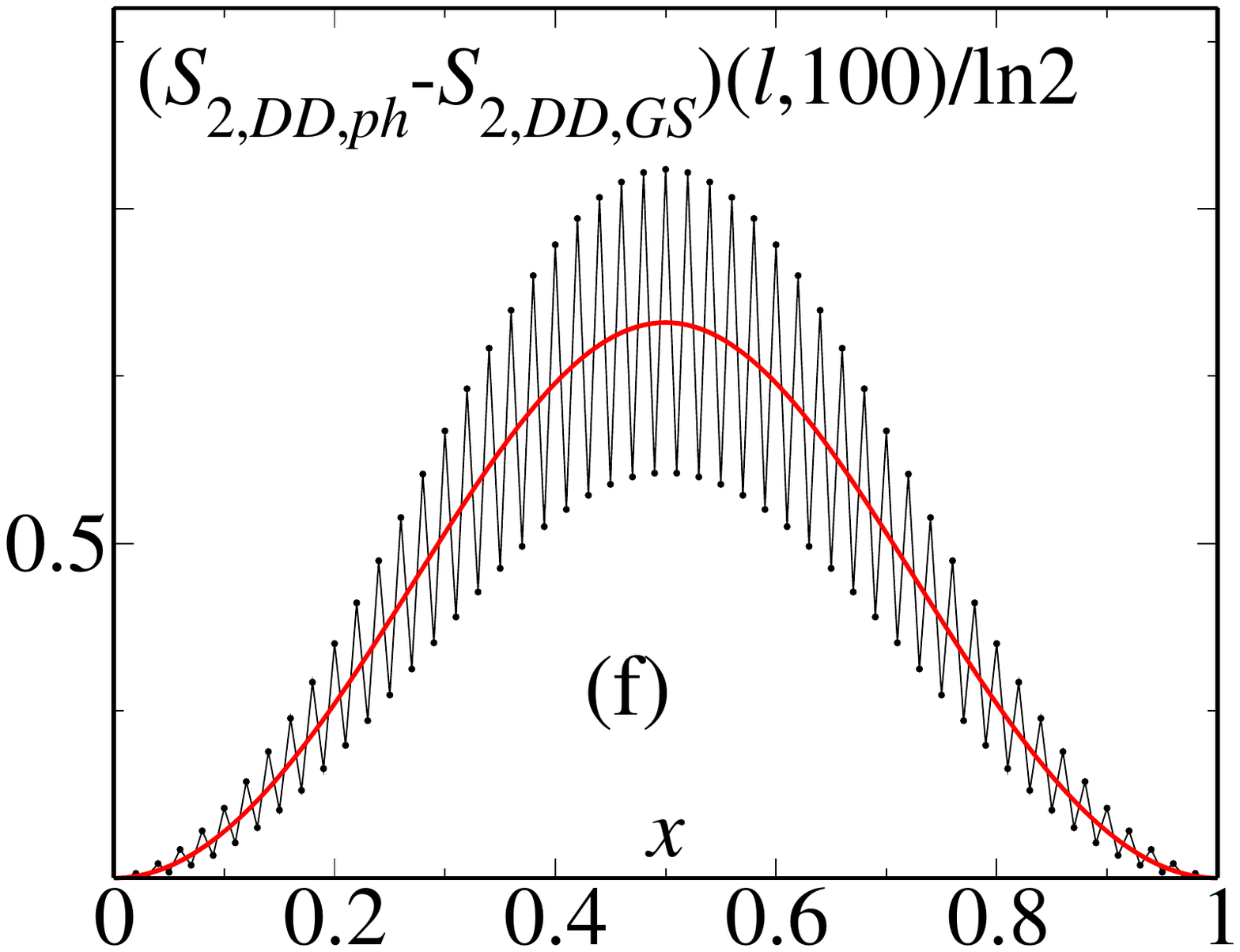}
  \includegraphics[width=0.49\textwidth]{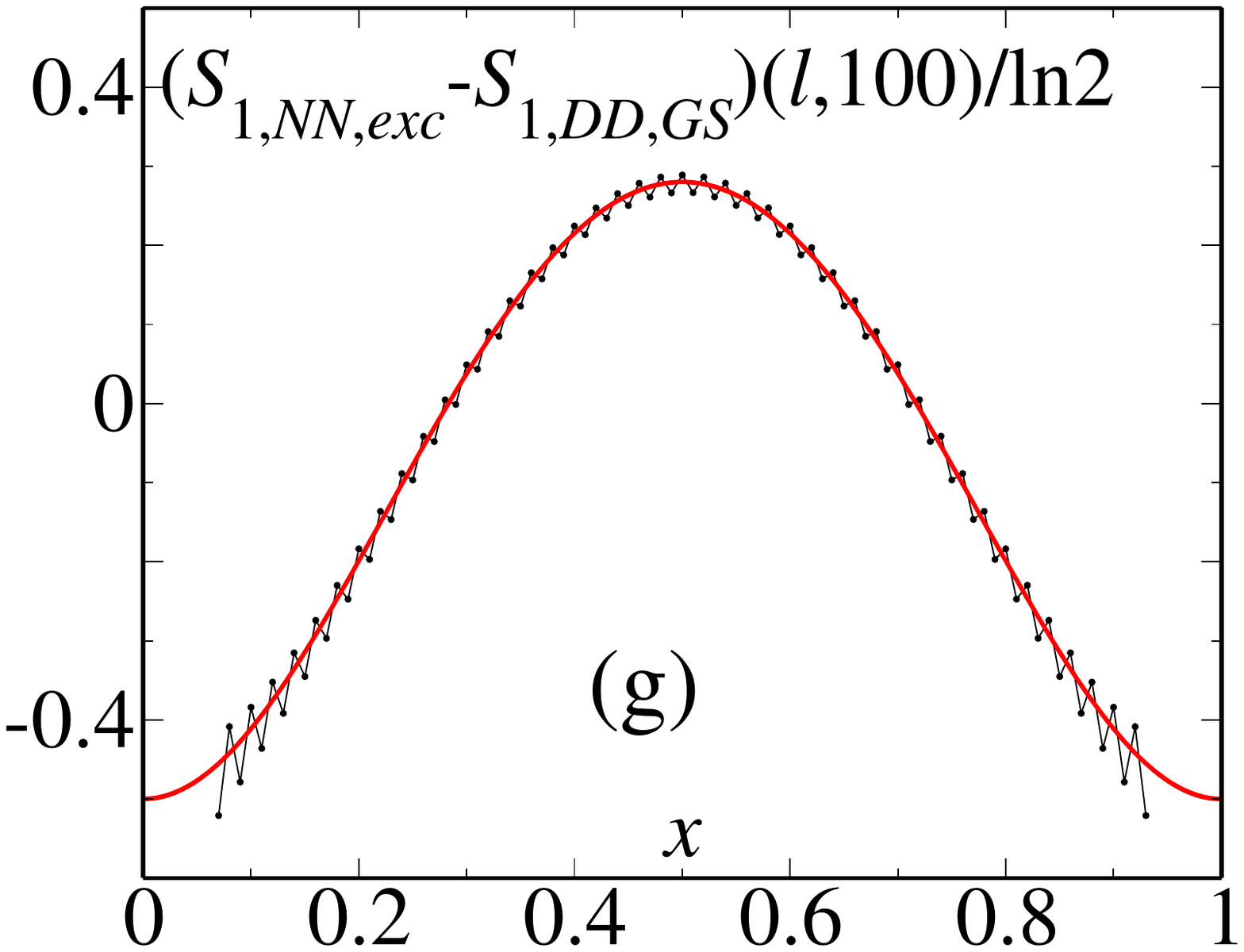}
  \includegraphics[width=0.49\textwidth]{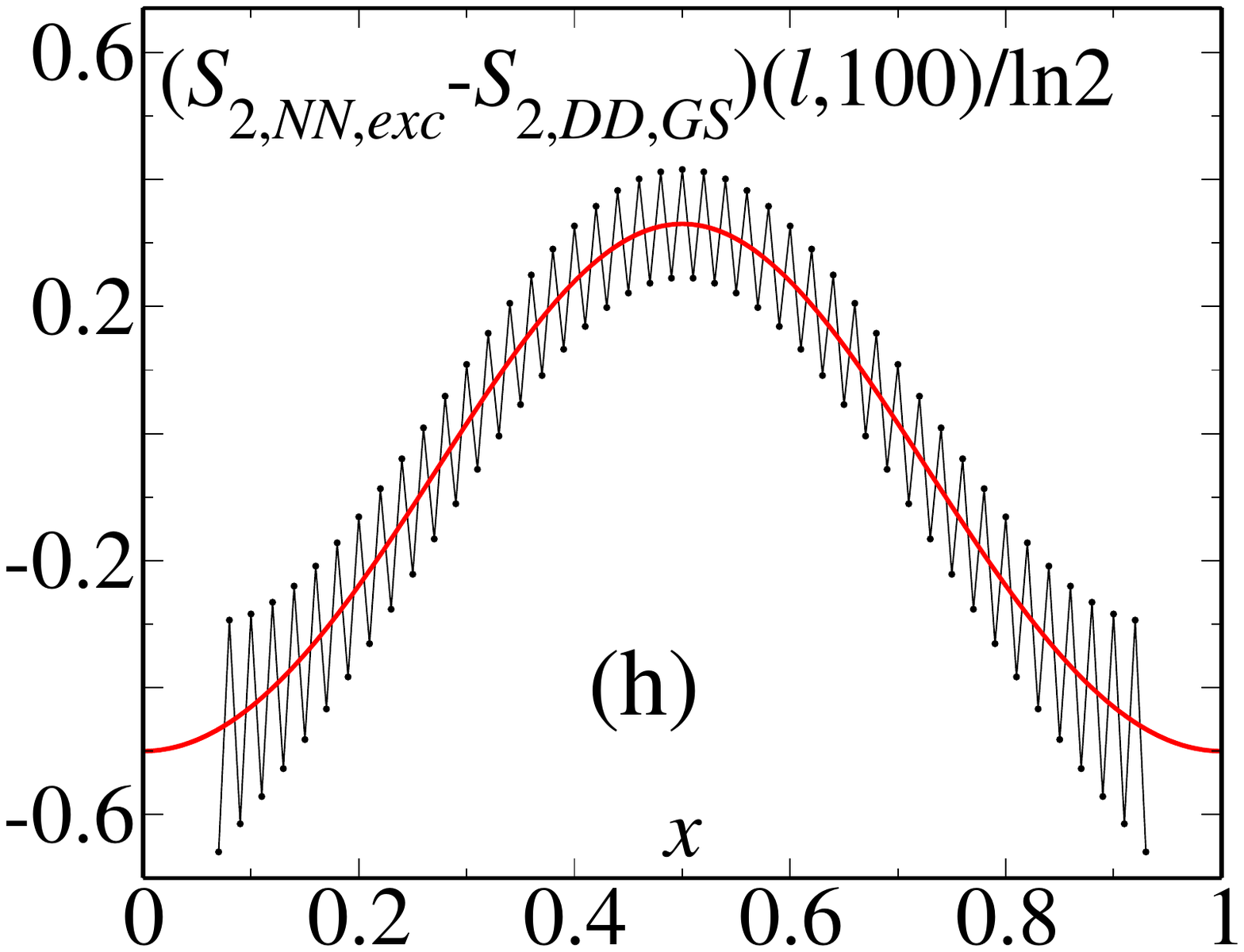}
 \end{minipage}
 \caption{REE's in the model (\ref{bilstein}) (I). Black dotted lines: $L=100$ numerical data; red solid line: CFT predictions (the BE's are added to the CFT formulas when necessary).}\label{XX_fig_1}
\end{figure}

The lattice realization of this CFT is given by a spin-1/2 XX chain, with  boundary couplings  \cite{Bilstein2000}
\begin{equation}\label{bilstein}
 \begin{split}
  H_B= & -\sum_{j=1}^{L-1}\left(\sigma_j^x\sigma_{j+1}^x+\sigma_j^x\sigma_{j+1}^x\right)+\\
  - & \frac{1}{2}(\alpha_-\sigma_1^-+\alpha_+\sigma_1^++\alpha_z\sigma_1^z+\beta_-\sigma_L^-+\beta_+\sigma_L^++\beta_z\sigma_L^z),
 \end{split}
\end{equation}
where the $D$BC, on a given edge,  is realized by setting all the boundary couplings 
to zero, while the $N$BC is realized, say  on the left side, by  choosing $\alpha_z=0$ 
and $\alpha_+=\alpha_-=2$ (moreover, in the sermonic picture of the  $DD$BC case, 
one has to work  in the half-filled sector). 
The operator content of the various models described by the Hamiltonian (\ref{bilstein}) 
can be expressed in terms of the partition functions of a free boson with different BC's
\cite{Bilstein2000,Saleur1998}. 
The $DD$ and $NN$ partition functions can be written in terms of the characters (\ref{Kharacters}) 
\barray
  Z_{DD}(q) &=& K_0(q)+K_2(q), \label{DDNN1} \\
  Z_{NN}(q) &= & K_0(q). \label{DDNN2}
\earray 
The absence of corrections for the GS in the $DD$BC case, with the exception of the usual oscillating ones \cite{Calabrese2010}, 
has already been observed  \cite{DalmonteErcolessiTaddia2011} and we confirm it numerically. 
 We expect the same feature for the GS in the $NN$BC case, which we analyze with DMRG 
 for system size $L=100$, keeping up to 1100 states, using 3 sweeps and achieving a truncation error of $10^{-10}$. 
The results are shown in  Figs. \ref{XX_fig_1}(a), (b), for $n=2, 3$ : up to oscillating corrections, typical of $c=1$ 
systems, and the BE $-\frac{1}{2}\ln 2$, as expected from Eq. (\ref{fcom}) we do not see any non-constant correction.

We then consider the first ES with $DD$BC  which, according to Eq. (\ref{DDNN1}), 
is generated by a vertex operator with  conformal dimension 1/2. The $F^{(n)}_\Upsilon$ functions
of  vertex operators $\Upsilon$ are equal to 1, for all $n$, so the  REE's receive no corrections \cite{AlcarazBerganzaSierra2011}. 
We verify this result by identifying this excitation in the fermionic version of the Hamiltonian (\ref{bilstein}) 
with the addition (or subtraction) of a fermion at half filling. 
The Figs. \ref{XX_fig_1}(c), (d) show the  $n=1,\,2$ REE's for these states, which confirmed the absence of corrections. 


For $DD$ and $NN$BC, there is another ES, with conformal weight $h=1$,
associated to the primary field $i\partial\phi$. For $DD$BC it corresponds to 
the lowest  particle-hole state created from the half-filled ground state \cite{AlcarazBerganzaSierra2011,BerganzaAlcarazSierra2012};
in the $NN$BC case, there is no particle number conservation, so it is simply the first ES. 
In the former  case, we use the method of \cite{Peschel2003}, and in the latter the  multi-target DMRG \cite{DegliEspostiBoschiOrtolani2004},
to compute  the relative REE's, shown in Figs \ref{XX_fig_1}(e), (f), (g), (h). Up to oscillations 
(and the BE  in the $NN$ case), we find excellent agreement with Eq. (\ref{ELC}).

\begin{figure}[t]
 \begin{minipage}{0.5\textwidth}
  \includegraphics[width=0.49\textwidth]{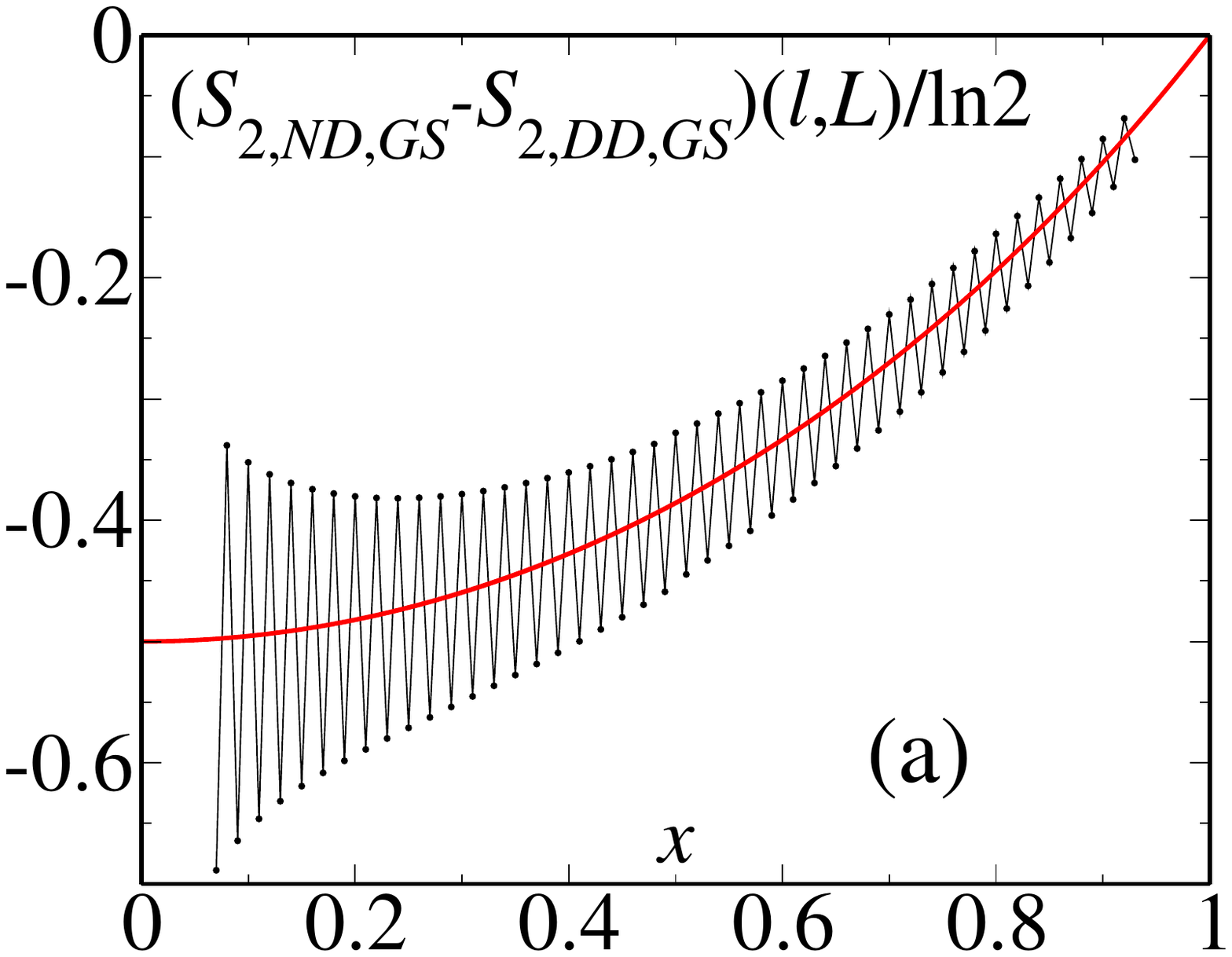}
  \includegraphics[width=0.49\textwidth]{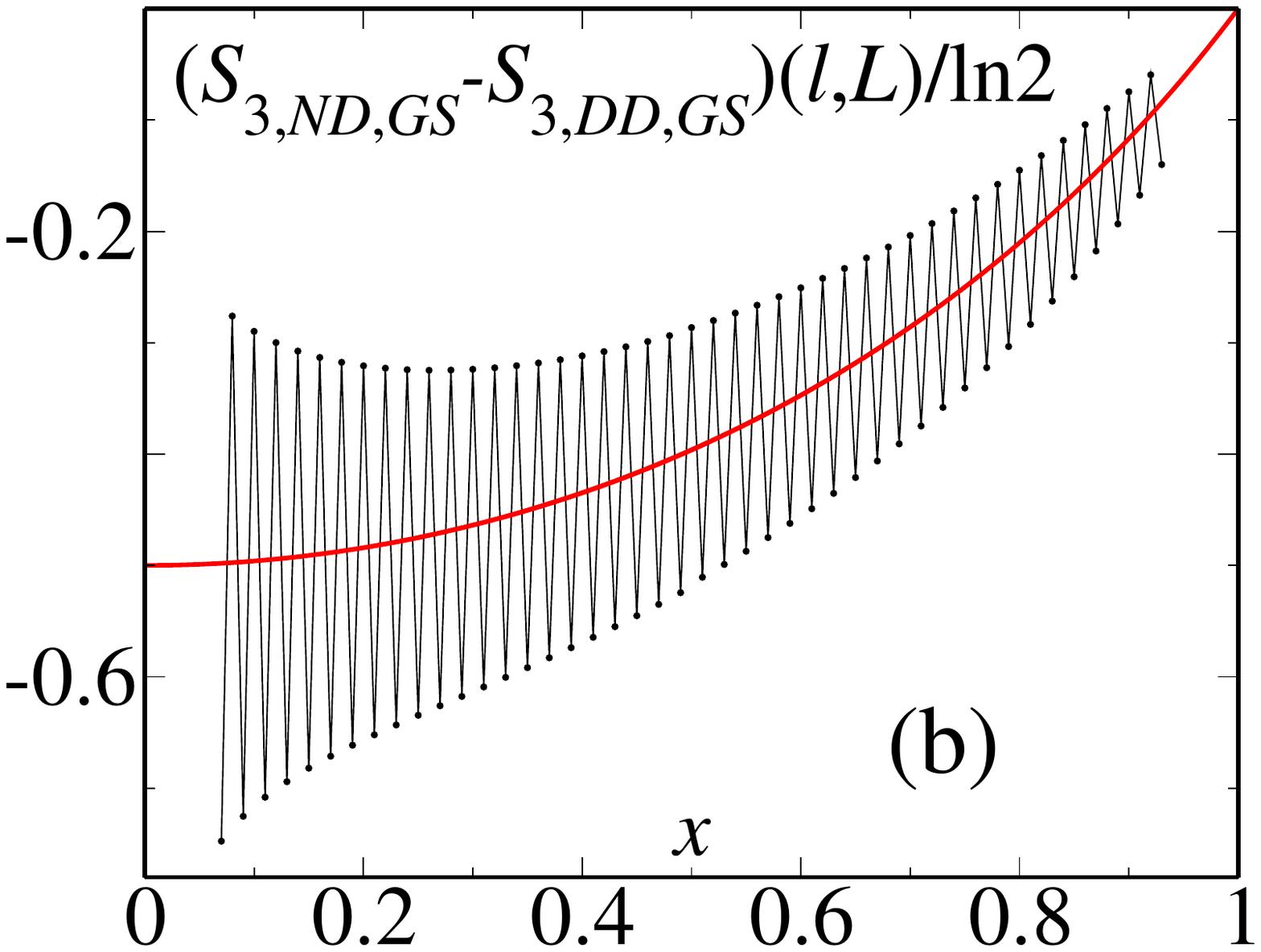}
  \includegraphics[width=0.49\textwidth]{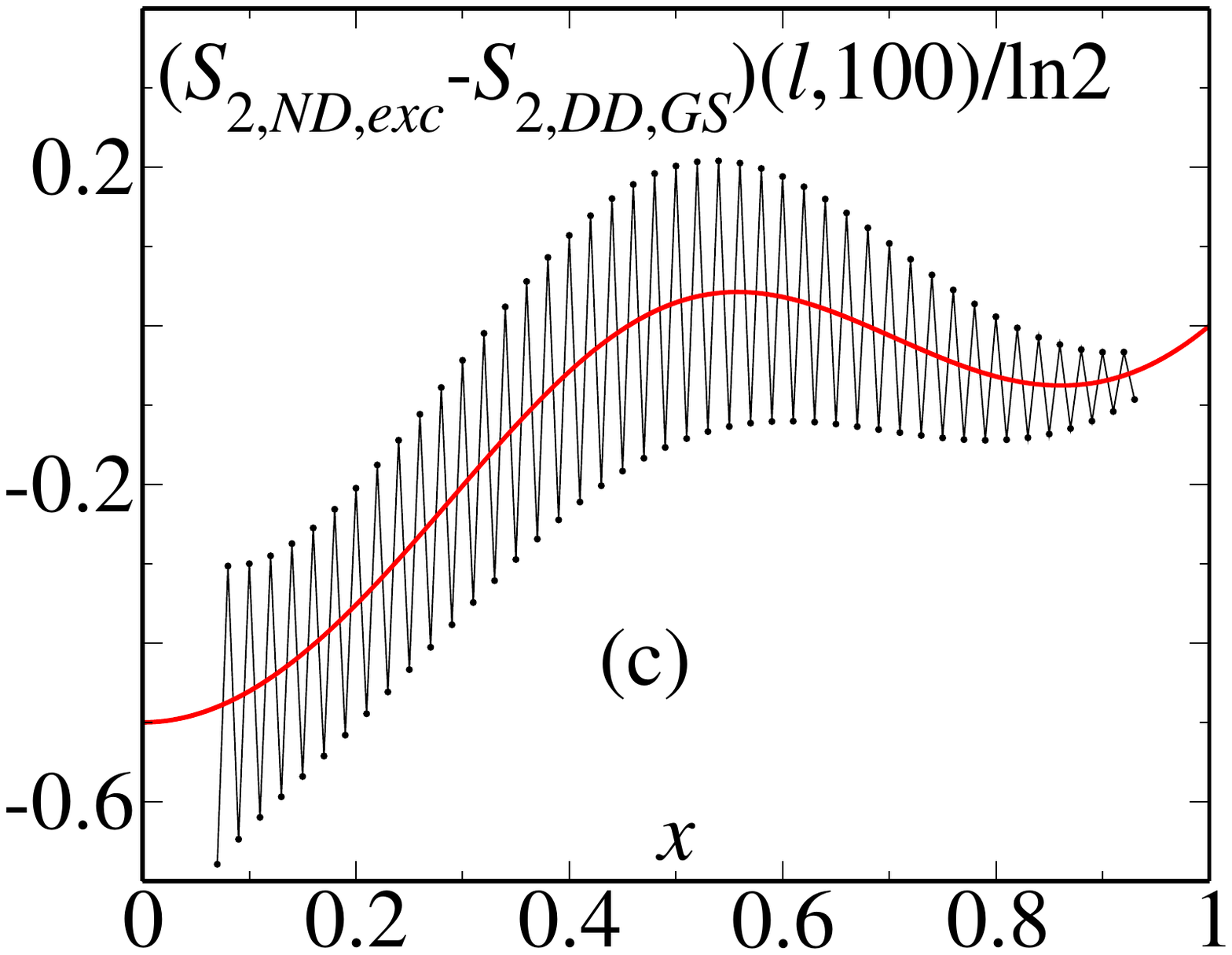}
  \includegraphics[width=0.49\textwidth]{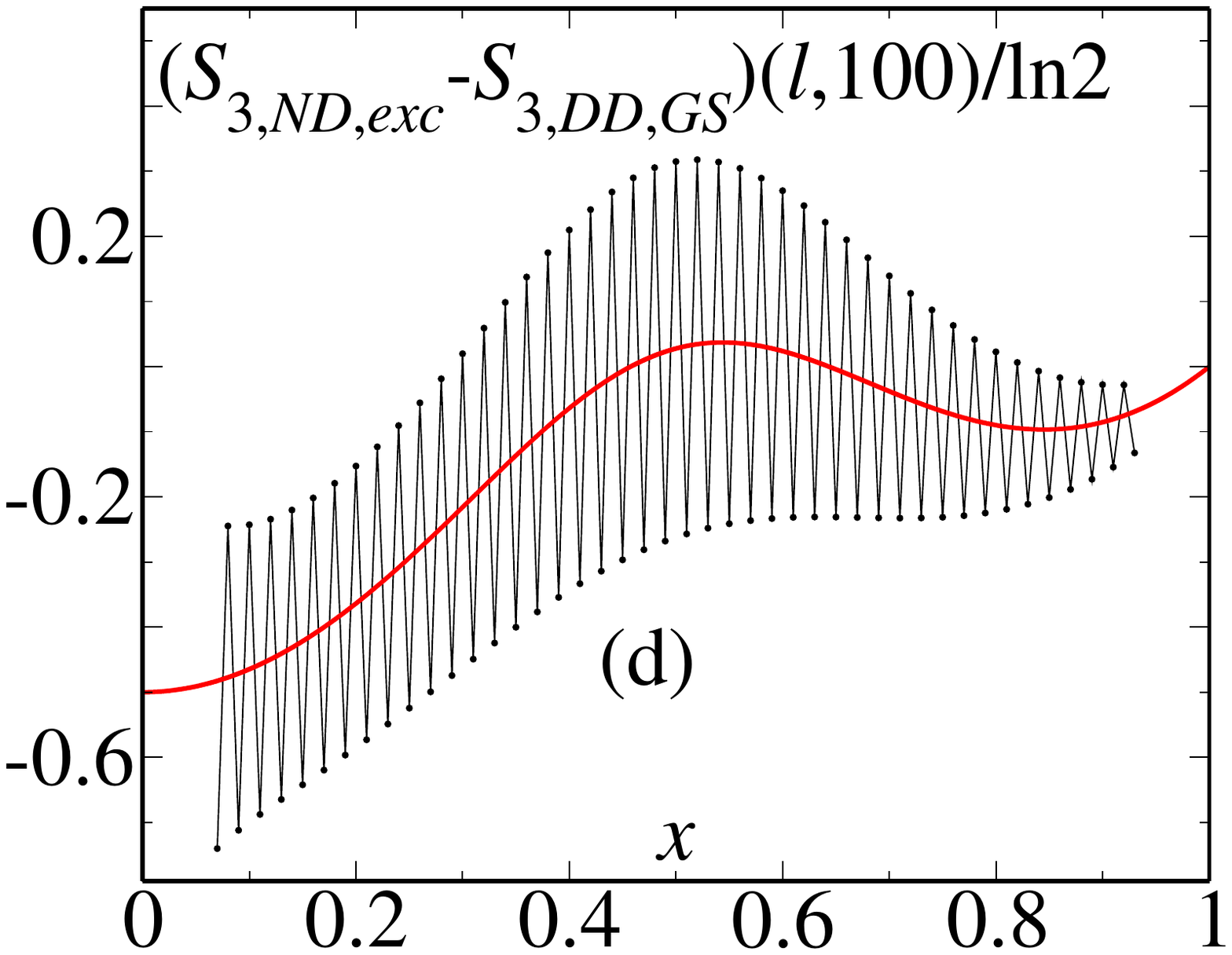}
 \end{minipage}
 \caption{REE's in the model (\ref{bilstein}) (II). See the caption of figure \ref{XX_fig_1}.}\label{XX_fig_2}
\end{figure}

Finally,  we study  the $ND$BC case, whose partition function \cite{Saleur1998} cannot be written in terms of the characters (\ref{Kharacters}). However, it can be shown  that
\begin{equation}
 Z_{ND}(q)=\chi_{1/16}(q)[\chi_0(q)+\chi_{1/2}(q)],
 \label{and}
\end{equation}
where $\chi_{0, 1/2, 1/16}$ are the characters of the primary fields  of the Ising  model. 
Equation (\ref{and})  implies that the operator content corresponding to $ND$BC  is given by 
the tensor product of two Ising models. 
In particular, the GS is  associated with the operator
$\mathbb{I}\otimes\sigma$, resulting in the correction $F_\sigma^{(n)}$, and the first ES with
$\sigma\otimes\chi$, resulting in the correction $F_\sigma^{(n)}F_\chi^{(n)}$
(plus a BE contribution $-\frac{1}{2}\ln 2$): we show in Fig.
\ref{XX_fig_2} the results for the $n=2,\,3$, obtained with multi-target DMRG, finding, even in this case, a remarkable agreement between CFT and numerics.

We conclude that for the critical Ising and XX quantum spin chains under any conformal BC's, the REE's of the low-lying states are obtained, apart from the BE's and the oscillations, from the correlators of the primary operators of the underlying CFT: the finite-size behavior of the REE's of quantum chains with open boundaries identify the primary operators in the CFT, providing, in principle, a tool for deepening the understanding of the operator content of a CFT. 



{\it Acknowledgements.} We thank M. I. Berganza, P. Calabrese, M. Dalmonte and E. Ercolessi for helpful discussions. 
L. T. thanks F. Ortolani for his unvaluable help with the DMRG code.
L. T. acknowledges financial support from  INFN COM4 grant NA41, 
J. C. X. and F. C. A.  from the Brazilian agencies FAPEMIG, FAPESP, and CNPq, and 
G. S. from the grants FIS2012-33642, QUITEMAD and the Severo Ochoa Program.

\end{document}